# Utilizing Automated Breast Cancer Detection to Identify Spatial Distributions of Tumor Infiltrating Lymphocytes in Invasive Breast Cancer

Han Le[1,+], Rajarsi Gupta[2,3,+], Le Hou[1], Shahira Abousamra[1], Danielle Fassler[3], Tahsin Kurc[2], Dimitris Samaras[1], Rebecca Batiste[3], Tianhao Zhao[3], Arvind Rao[4], Alison L. Van Dyke[5], Ashish Sharma[6], Erich Bremer[2], Jonas S. Almeida[7], Joel Saltz[2,*]

*From Department of Computer Science, Stony Brook University, Stony Brook, NY,[1] Department of Biomedical Informatics, Stony Brook Medicine, Stony Brook, NY,[2] Department of Pathology, Stony Brook University Hospital, Stony Brook, NY,[3] Department of Computational Medicine & Bioinformatics, University of Michigan Medical School, Ann Arbor, MI,[4] Surveillance Research Program, Division of Cancer Control and Population Sciences, National Cancer Institute, National Institutes of Health, Bethesda, MD,[5] Department of Biomedical Informatics, Emory University, Atlanta, GA[6] National Cancer Institute Division of Cancer Epidemiology & Genetics, Bethesda, MD,[7] these authors contributed equally to this work,[+] corresponding author: joel.saltz@stonybrookmedicine.edu,[*]*

Quantitative assessment of Tumor-TIL spatial relationships is increasingly important in both basic science and clinical aspects of breast cancer research. We have developed and evaluated convolutional neural network (CNN) analysis pipelines to generate combined maps of cancer regions and tumor infiltrating lymphocytes (TILs) in routine diagnostic breast cancer whole slide tissue images (WSIs). The combined maps provide 1) insight about the structural patterns and spatial distribution of lymphocytic infiltrates and 2) facilitate improved quantification of TILs. We evaluated both tumor and TIL analyses using three CNN networks - Resnet-34, VGG16 and Inception v4, and demonstrated that the results compared favorably to those obtained by the best published methods. We have produced open-source tools and a public dataset consisting of tumor/TIL maps for 1,015 TCGA invasive breast cancer images. The maps can be downloaded for further downstream analyses.

Among women worldwide, invasive breast cancer is the most common cancer and the second most common cause of cancer-related deaths [1], despite decreasing mortality rates in recent years due to early diagnosis and current therapeutic options that significantly prolong survival. Invasive breast cancers are a heterogeneous category of disease phenotypes [2], [3] that are histologically classified into subtypes based on growth patterns; the expression of estrogen (ER), progesterone (PR), human epidermal growth factor receptor 2 (HER2); and the Ki-67 proliferation index.

The role of tumor infiltrating lymphocytes (TILs) in invasive breast cancer has become increasingly important as a biomarker that can predict clinical outcomes, as well as treatment response in the neoadjuvant and adjuvant setting [4], [5], [6], [7],[8], [9], [10], [11]. TILs are a readily available biomarker and their evaluation is likely to expand with the emergence of immunotherapy. Elevated concentrations of TILs in HER2-positive [12] and triple-negative (ER-/PR-/HER2-) [13] breast cancers are associated with prolonged overall and disease-free survival; whereas elevated concentrations of TILs in luminal HER2-negative breast cancer have been associated with poor overall survival [4]. TILs can also serve as a predictive biomarker since a significant part of the cytotoxic effects of systemic chemotherapy and radiation therapy are actually mediated by activating the immune system to kill cancer cells instead of directly targeting the tumor cells [14]. Targeted therapies against HER2 and vascular endothelial growth factor (VEGF) are mediated by both antibody-dependent and complement-mediated cytotoxicity in cancer cells through lymphocytes and other immune cells in the tumor microenvironment [15]. Recent studies suggest potential for synergistic effects between targeted and immune therapies in multiple disease sites [16], [17].

Current practice routinely includes manual assessments of hematoxylin and eosin (H&E) stained tissue sections by surgical pathologists to identify and classify invasive breast cancer. Such diagnostic evaluation provides insight about clinical management, treatment selection, survival, and recurrence. Since H&E tissue sections are readily available, there is a sustainable opportunity to provide potentially actionable data about TILs without the need for additional tissue samples -- e.g., immunohistochemical (IHC) testing. H&E tissue also permits the interpretation of the lymphocyte infiltrate within and proximal to the tumor in the context of histology to provide insight about the spatial relationships between tumor regions and TILs. The published guidelines for the histologic assessment of TILs in invasive breast cancer [18], [19], [20] require pathologists to select the region of tumor and to delineate stromal areas in order to assess the percentage of TILs (%TILs) in *stromal* regions as a continuous variable from 0-100% within the boundaries of the entire tumor that is used to classify the lymphocyte infiltrate as low, intermediate, and high, respectively. However, this evaluation is intrinsically qualitative and often subject to inter-observer variability, so previous work has articulated these concerns [21] in an attempt to clearly state the need for automated methods to evaluate %TILs in H&E tissue sections of breast cancer. Computationally calculating %TILs intrinsically provides spatial information about how TILs are distributed in whole slide images (WSIs), where it is likely that the distinction between intratumoral and stromal TIL infiltrates is important. While there have been some relatively small studies examining intratumoral and stromal TILs [22], the predictive power of the spatial distribution of TILs within tumor and tumor-associated stroma needs to be better elucidated. Automated evaluation of TILs in H&E WSIs fundamentally requires tumor segmentation linked with the detection of lymphocyte infiltrates. Automation of H&E tumor-TIL analyses will make it possible to carry out large-scale correlative studies that quantitatively characterize TIL distributions in well-characterized clinical populations. Computer analysis of high-resolution images of whole slide tissue specimens can enable a data driven and quantitative characterization of TIL patterns.

With the recent success of deep learning [23] and the availability of public datasets [24], [25], [26], [27], several research groups have proposed deep learning based algorithms to detect or segment cancer/tumor regions in breast cancer WSIs [28], [29], [30], [31]. Previous methods developed classification models from customized convolutional neural networks [28], [29] or from limited training data [30], [31].

In our work, we use standard state-of-the-art deep learning models along with a large-scale dataset to detect invasive breast cancer regions in WSIs. Our approach automates breast cancer detection at intermediate- to high-resolution in order to generate detailed probability-based heatmaps of the tumor bed. It achieves an F1-score of 0.82, a positive predictive value (PPV) of 79%, and negative predictive value (NPV) of 98% in terms of pixel-by-pixel evaluation in an unseen and independent test dataset consisting of 195 WSIs from the Cancer Genome Atlas (TCGA) repository. These performance numbers are better than those achieved by the models in the previous works.

Moreover, our study combines tumor detection with lymphocyte detection to identify tumor-TIL patterns in a large number of publicly accessible WSIs. We trained TIL prediction models using training datasets from a previously published deep learning approach [32] to generate high resolution TIL maps. We then combined the cancer detection results with the TIL results. The combined results represent regions of tumor with intra- and peri-tumoral TILs in publicly available 1,015 WSIs from the TCGA repository. We expect that the availability of high-resolution spatial Tumor-TIL maps will allow quantitative estimation and characterization of the relationship between tumor cells and TILs. The ability to quantify and visualize the spatial relationships between tumor and TILs can be a very practical and useful way to further elucidate intriguing observations in previous studies. It will also further our collective understanding of the biological behavior of invasive breast cancers within the context of cancer-immune interactions in the tumor microenvironment.

**MATERIALS AND METHODS**

**Ethics Statement**

We used high-resolution WSIs from the Surveillance, Epidemiology, and End Results (SEER: https://seer.cancer.gov/) cancer registry system and from TCGA (https://portal.gdc.cancer.gov/) to train and evaluate the deep learning models and generate cancer region maps. The WSIs from TCGA are de-identified and publicly available for research use. The WSIs from SEER came from a pilot program examining the feasibility of and best practices for a Virtual Tissue Repository (VTR Pilot). As all data in the VTR Pilot, including the whole slide images, had been deidentified prior to receipt, the NIH Office of Human Subjects Research Protection determined that the study was excluded from NIH IRB review. Each of the SEER registries supplying the deidentified WSIs has obtained IRB approval from their respective institution(s). The Stony Brook IRB has classified the dataset as being a non-human subjects research dataset.

**Datasets**

The training, validation, and test datasets for the breast cancer detection models consisted of image patches extracted from 102, 7, and 89 SEER WSIs, respectively. All of the images were scanned at 40x magnification and manually segmented by an expert pathologist into cancer and non-cancer regions using a web-based application [33]. Additionally, we evaluated the deep learning models with 195 TCGA WSIs (referred to here as $T_{tcga}$), which had been manually annotated in work done by Cruz-Roa et al. [29]. The details of the training, validation and test datasets for tumor region segmentation are presented in Table 1. The trained models were applied to 1,015 diagnostic WSIs from TCGA invasive breast cancer cases.

The same set of 1,015 WSIs was also analyzed using the TIL classification models trained with data generated by Saltz et al. [32]. These data consisted of 86,154 and 653 image patches for training and validation, respectively. We created a test dataset of 327 patches extracted from TCGA invasive breast cancer WSIs to evaluate the trained TIL models. The details of the training, validation, and test datasets for TIL classification are presented in Table 2.

| Source | Purpose | ID | WSIs (N) | Patches (N) | Cancer-positive (N) | Cancer-negative (N) |
|---|---|---|---|---|---|---|
| SEER | Training | $D_{tr}$ | 102 | 333,604 | 99,889 | 233,715 |
| | Validation | $D_{val}$ | 7 | 10,224 | 4,953 | 5,271 |
| | Testing | $T_{seer}$ | 89 | - | - | - |
| TCGA | Testing | $T_{tcga}$ | 195 | - | - | - |

TABLE 1
Data statistics of the training, validation and testing datasets for the breast cancer detection models.

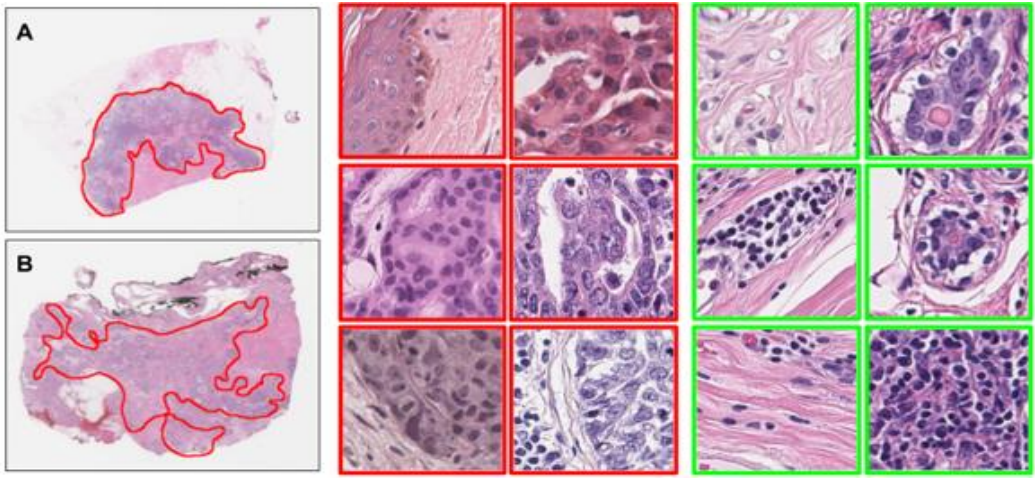

Fig. 1. Annotation example from pathologist (Figures A and B) and image patches extracted from WSIs. Red lines in figures A and B were manually drawn by expert pathologist to enclose the cancer regions. Regions outside the annotated regions are non-cancer regions. Patches surrounded in red boxes are positive samples which contain invasive cancer cells. Patches surrounded in green boxes are negative samples which do not contain invasive cancer cells.

| Srouce | Purpose | Patches (N) | TIL-positive | TIL-negative |
|---|---|---|---|---|
| *TCGA* | Training | 86,154 | 21,773 | 64,381 |
| | Validation | 653 | 295 | 357 |
| | Testing | 327 | 174 | 153 |

TABLE 2
Data statistics of the Lymphocytes dataset provided in Saltz et al. [32]

**Patch Extraction for Breast Cancer Detection Models**

We extracted image patches at the highest image resolution within and outside the manually segmented cancer regions using an open source library called OpenSlide [34]. Patches with a size of 350 x 350 pixels at 40x magnification (equivalent to 88µm x 88µm) resulted in the best classification performance and were used to create the training datasets. Each patch was labeled cancer-positive (i.e., it intersected or was in a cancer/tumor region) or cancer-negative (i.e., it was outside cancer/tumor regions). Figure 1 shows an example of the pathologist's annotations. The region inside the red line represents the cancer region. The figure also shows the sample patches extracted from the cancer and non-cancer regions.

Previous work has shown that it is beneficial to have more negative samples than positive samples in a training dataset for image classification in digital pathology [29], [35], [36], [37]. A good ratio of negative to positive patches will increase the generalization of a convolutional neural network (CNN) model and decrease false positive rate. We experimented with a range of ratios of cancer-negative patches to cancer-positive patches with the same validation dataset. The final training, validation, and test datasets are presented in Table 1.

**Convolutional Neural Networks**

We investigated and adapted multiple state-of-the-art deep learning architectures, namely the VGG 16-layer [38], the Resnet 34-layer [39], and the Inception-v4 network [40]. These are state-of-the-art CNNs which are widely used in a wide range of application domains. VGG16 and Resnet34 are designed to process 224x224-pixel patches. Inception-v4 accepts 299x299-pixel image patches. Our tumor dataset consists of 350x350-pixel patches at 40x magnification and the Lymphocyte training datasets contain 100x100-pixel patches at 20x magnification. Input patches in these datasets were resized to the desired input size for each network. In addition, for Resnet34 and Inception-v4, we changed the dimension of the output layer from 1,000 classes to two classes, because each patch in our case is labeled positive or negative.

| VGG16 | | Resnet34 | | Inception-v4 | |
|---|---|---|---|---|---|
| Original | Modified | Original | Modified | Original | Modified |
| Linear(25088, 4096) ReLU → Dropout Linear(4096, 4096) ReLU → Dropout Linear(4096, 1000) | Linear(25088, 4096) ReLU → Dropout Linear(1024,2) | Linear(512,1000) | Linear(512,2) | Linear(1536,1000) | Linear(1536,2) |

TABLE 3
Modifications to the classification layers of the CNNs

For VGG16, we reduced the size of the intermediate features of the classification layer from 4,096 to 1,024 and only kept the first four layers in the classification layer. This modification reduced the number of trainable parameters of this network from 138 million to 41 million. Our modifications to the classification layers of the CNN architectures are presented in Table 3. We implemented the CNN networks using pyTorch 0.4 [44].

Earlier work [41], [42] showed that refining a CNN pre-trained on the ImageNet dataset [43] is a good approach to boosting image classification performance in digital pathology. We refined the pre-trained CNN models with our training data. We used the same training procedure for all of the networks. At the beginning of the training, the weights of the networks were initially fixed except for the classification layer. The networks were trained in this state for N epochs (N is three for the Cancer models and N is five for the Lymphocyte models) with a batch size of B (B is 256 for the Cancer models and B is 128 for the Lymphocyte models), an initial learning rate of 0.01, a momentum of 0.9 and a weight decay of 0.0001. After N epochs, the training process turned on updates to the initially fixed weights. The network was then trained for total of 20 epochs, updating all the weights. The training process used a stochastic gradient descent method [45] in order to minimize a cross entropy loss function.

The color profiles of WSIs may vary from image to image because of variations in staining and image acquisition [46], [47],[48]. We normalized the R, G, and B channels of each patch to a mean of 0.0 and standard deviation of 1.0. Additionally, we employed data augmentation to further reduce the effects of color/intensity variability and data acquisition artifacts. The data augmentation operations included random rotation between 0 and 22.5 degrees, random vertical and horizontal flipping, perturbations in patch brightness, contrast, and saturation. In the prediction (test) phase, no data augmentation was applied except for the normalization of the color channels. Each patch was assigned a value between 0.0 and 1.0 by the trained model, indicating the probability of the patch being positive.

**Experiments**

In our experimental evaluation, we used accuracy, F1-score, and area under the ROC (Receiver Operating Characteristic) curve (AUC) as performance metrics. Accuracy is the ratio of correctly classified patches to the total number of patches in the ground truth test dataset. Because a dataset is not always balanced between classes, we used the F1-score that considers both precision and recall to compute a score. Mathematically, F1-score is equal to 2*(precision*recall)/(precision + recall). Lastly, we used AUC to evaluate the prediction performance of the models at different threshold settings. AUC shows the relationship between the True Positive Rate (TPR) and False Positive Rate (FPR) of a model. It is a widely used metric to assess model performance for binary classification tasks.

Tables 4 and 5 show the cancer region segmentation and TIL classification performances of the different CNNs, respectively (please see the Results section for details). The best models were applied to the 1,015 WSIs from TCGA invasive breast cancer cases to generate what we call *prediction probability maps* for cancer regions and TILs. A prediction probability map is constructed by uniformly partitioning a WSI into image patches in each image dimension. The image patches are analyzed by a trained model and assigned a label probability between 0.0 and 1.0. For cancer region segmentation, the label of a patch was either cancer-positive (i.e., the patch predicted to be within or intersect a cancer region) or cancer-negative (i.e., the patch is predicted to be outside the cancer regions in the WSI). For TIL classification, the label of a patch was either TIL-positive (i.e., the patch was predicted to contain lymphocytes) or TIL- negative. We implemented a Web-based application to visualize and interact with the prediction probability maps as heatmaps (please see Figure 2 and the Methods section).

**Post-processing Step for Cancer Heatmaps**

Most patch-based classification algorithms [49], [50] predict the label of a patch independent of other patches in an image. They do not take into account the characteristics and labels of neighbor patches. Invasive cancer regions in breast cancer tend to be close to each other. In other words, the probability of a patch to be cancer-positive is correlated with its surrounding patches. To incorporate this information in our analysis pipeline, we employed a simple, yet effective, aggregation approach as a post-processing step. This approach takes per-patch classification probability values, converts them into a probability map, called *H*, and produces an aggregated probability map, called *A*. The classification probability value of a patch in *A* is computed by an aggregation operation over neighbor patches within a specific distance of the patch in *H*. The relationship between *A* and *H* can be formulated as follows:

$$A(i,j) = f\left(\left\{H(m,n) \mid m,n \in \left[\left\lfloor\frac{i}{w}\right\rfloor w, \left(\left\lfloor\frac{i}{w}\right\rfloor+1\right)w\right]\right\}\right)$$

Here, *H*(m,n) is the probability values of a patch at location (m,n) in *H*. *A*(i, j) is the probability value of the aggregated patch at location (i,j) in *A*. f is the aggregation function over a set of patches in a window of

$$\left[\left\lfloor\frac{i}{w}\right\rfloor w, \left(\left\lfloor\frac{i}{w}\right\rfloor+1\right)w\right] \subset \left[\left\lfloor\frac{i}{w}\right\rfloor w, \left(\left\lfloor\frac{i}{w}\right\rfloor+1\right)w\right]$$

and w is the window size. In our aggregation approach, all patches within the window will have the same prediction score after the aggregation operation. $\lfloor x \rfloor$ is the Floor operation which takes x as an input and returns the largest integer that is less than or equal to x. We explored different aggregation function such as Average, Median, and Max. The experiments were carried out using $T_{seer}$. The best aggregation method from these experiments was used to generate aggregated probability maps for $T_{tcga}$. Empirically, we found that the Max function and a window of 4x4 resulted in the best performance with $T_{seer}$. We applied these settings to post-process predictions in $T_{tcga}$.

**Combined Tumor-TIL Maps**

We merged each pair of cancer and lymphocyte heatmaps into a single heatmap as an RGB image. The R channel stores the lymphocyte probabilities quantized to 0-255; the G channel stores the cancer probabilities quantized to 0-255; and the B channel stores 0 or 255 to indicate if a patch is glass background or tissue, respectively.

**Software Support for Analysis Workflow**

We have employed software called QuIP [51] and caMicroscope [33] to support the data management and visualization requirements in our study. A typical whole slide tissue image can be several Gigabytes (GBs) in size. Even a modest cohort of a hundred subjects can result in one Terabyte (TB) of image data. It is a non-trivial task to efficiently store, manage and index a dataset of this size and to provide interactive capabilities for visualization of images and analysis results for evaluation, validation, and additional downstream analyses. Examination of the analysis results, i.e., the probability maps, requires their interactive interrogation through visual analytic tools that link the probability maps with the underlying images. Our software converts probability maps into heatmaps for visualization purposes. We have developed a web-based application, called FeatureMap, and a database, called PathDB, in QuIP. PathDB manages and indexes metadata about whole slide tissue images and metadata about heatmaps. It links the heatmaps with the images for query and retrieval. FeatureMap implements a browser-based multivariate visualization library that is sufficiently lightweight to run on a mobile device. It interacts with PathDB to query and retrieve heatmaps, display them as low-resolution images so that a user can rapidly go through multiple images and probability maps. The low-resolution image representations of the probability maps are linked to full-resolution images and high-resolution heatmaps. The user can switch to the high-resolution view for more detailed and interactive examination of a probability map and the source image.

## RESULTS
### Evaluation of Cancer Detection Models

We trained three cancer detection and segmentation models, C-VGG16, C-Resnet34 and C-Incepv4, by using VGG16, Resnet34 and Inception-v4, respectively. We compared the performances of the models to each other as well as to another network, called ConvNet, which was developed by Cruz-Roa et al. [28], [29]. ConvNet was trained on a different training dataset, called HUP and UHCMC/CWRU [28], in the previous work. In order to use our training datasets, we implemented ConvNet using Pytorch [44] by precisely following the network description in the original paper [28]. We call our implementation ConvNet-ours.

We computed an average F1-score across all the test images by varying the threshold value from 0.0 to 1.0 in steps of 0.01. At each threshold value, prediction probability maps were computed for the 195 test images by the model under evaluation. The patch labels were assigned by applying the threshold value to the corresponding probability maps. The label maps and the ground-truth masks [29] were then used to compute average F1- score, positive predictive value (PPV), negative predictive value (NPV), true positive rate (TPR), true negative rate (TNR), false positive rate (FPR), and false negative rate (FNR). Table 3 shows the performance comparison between our models, the original ConvNet model [28], [29], and our implementation of the ConvNet model (ConvNet-ours). We report the performance of ConvNet- ours both with and without applying our post-processing step (see the Methods section), because the original ConvNet model did not include a post-processing step. Table 4 shows that the post-processing step improves the average F1-score from 0.75 to 0.77 and PPV from 0.69 to 0.73. Moreover, ConvNet-ours slightly outperforms the original ConvNet model in all metrics.

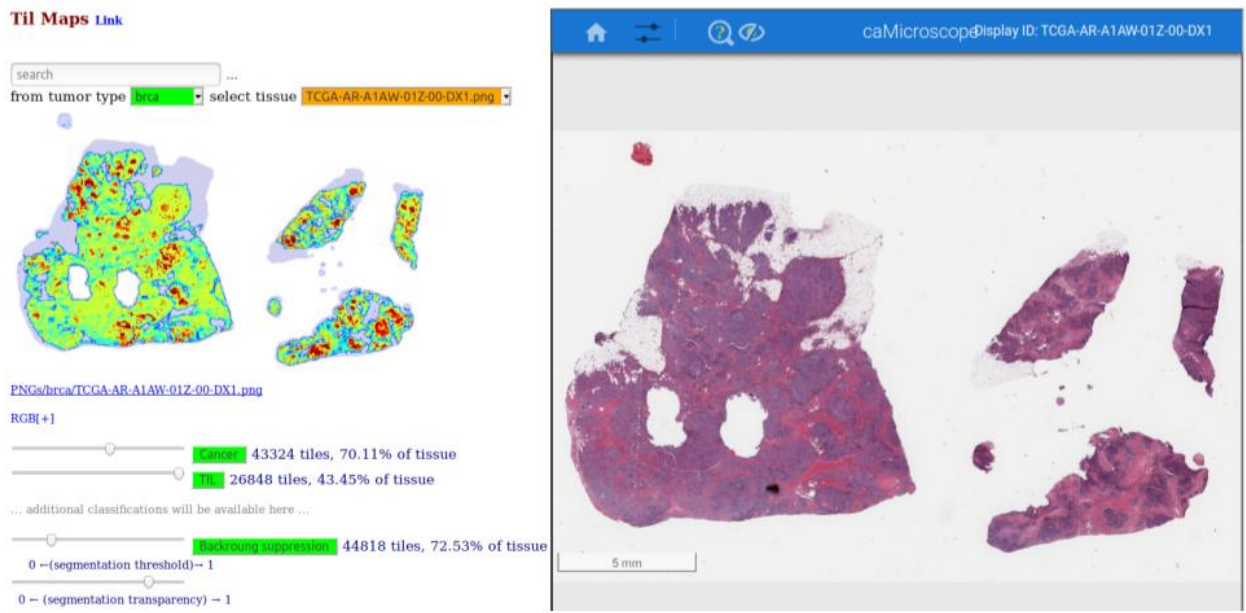

Fig. 2. User interface of our web-based application to study the spatial relationship between cancer regions and lymphocyte regions. Figure on the left shows the TILs heatmap where invasive breast cancer detection denoted in yellow with superimposed lymphocyte detection denoted in red. Figure on the right is the caMicroscope [33] that displays the regions of the WSI. Users can click on the TILs map to zoom in the corresponding regions on the caMicroscope.

| Method | F1-score | PPV | NPV | TPR | TNR | FPR | FNR |
|---|---|---|---|---|---|---|---|
| ConvNet [29] | 0.76 ± 0.20 | 0.72 ± 0.22 | 0.97 ± 0.05 | 0.87 ± 0.16 | 0.92 ± 0.08 | 0.08 ± 0.08 | 0.13 ± 0.16 |
| ConvNet-ours | 0.75 ± 0.18 | 0.69 ± 0.22 | 0.96 ± 0.09 | 0.87 ± 0.18 | 0.91 ± 0.09 | 0.09 ± 0.07 | 0.12 ± 0.16 |
| ConvNet-ours* | 0.77 ± 0.21 | 0.73 ± 0.23 | 0.97 ± 0.09 | 0.87 ± 0.23 | 0.92 ± 0.09 | 0.08 ± 0.09 | 0.13 ± 0.22 |
| C-VGG16 | 0.80 ± 0.20 | 0.78 ± 0.20 | 0.97 ± 0.05 | 0.88 ± 0.21 | 0.94 ± 0.06 | 0.06 ± 0.06 | 0.12 ± 0.21 |
| C-Resnet34 | **0.82 ± 0.18** | **0.79 ± 0.20** | **0.98 ± 0.04** | **0.89 ± 0.18** | **0.95 ± 0.05** | **0.05 ± 0.05** | **0.11 ± 0.18** |
| C-Incepv4 | 0.81 ± 0.19 | 0.79 ± 0.20 | 0.97 ± 0.05 | 0.88 ± 0.19 | 0.94 ± 0.06 | 0.06 ± 0.06 | 0.12 ± 0.19 |

TABLE 4

Performance comparison of the Cancer Detection task between the ConvNet [29] and our models. ConvNet-ours: Our implementation of the ConvNet [29] that was trained on the SEER dataset. The ConvNet-ours results are reported without applying the post-processing method (please see the Methods section for a description of the post-processing method). ConvNet-ours∗: Our implemented version of the ConvNet [29] that was trained on SEER dataset. The ConvNet-our* results are reported after the post-processing step is executed. The last three rows show the performances of our CNNs. All the models were trained on the SEER dataset ($D_{tr}$) and evaluated on 195 TCGA WSIs ($T_{tcga}$).

Figure 3 shows probability maps from the C-Resnet34 model for a set of representative WSIs in $T_{tcga}$. The shades of Red in the map images indicate the probability of a patch being cancer-positive as predicted by the model. Visual inspection of the maps and the respective WSIs showed that the model was able to detect and segment cancer regions well.

**Evaluation of Lymphocyte Classification Models**

We trained three lymphocyte detection models: L-VGG16, L-Resnet34, and L-Incepv4, using VGG16, Resnet34, and Inception-v4, respectively. We created a training dataset, containing 2,912 image patches from invasive breast cancer WSIs only, from the original TIL training dataset in work done by Saltz et al. [32]. The 86,154 patches in the original training dataset had been selected from multiple cancer types. Our experiments showed that the smaller training dataset resulted in more accurate classification models than the full original dataset.

We tested the trained models with a set of image patches extracted from TCGA invasive breast cancer WSIs. Table 5 shows the performance comparison between our models with the model developed in the previous work [32]. Our new models consistently outperformed the previous model in all of the performance metrics.

Our experimental evaluation showed that the cancer region segmentation and lymphocyte classification models achieved very good performance with respect to the F1-score, accuracy, and AUC metrics and performed better than the previous models. We applied the best of these models to 1,015 TCGA invasive breast cancer WSIs and generated Tumor, TIL, and combined Tumor-TIL maps. We will make these maps publicly available (see the Data Availability section). Figure 5 shows example Tumor-TIL combined maps overlaid on WSIs as heatmaps. The figure visualizes the spatial relationships between lymphocytes and tumor regions. The lymphocyte patches in these examples show TILs and tumor-associated lymphocytes (TALs) that surround the cancer regions. These visual representations of TILs, TALs, and cancer regions provide valuable information for further analyses.

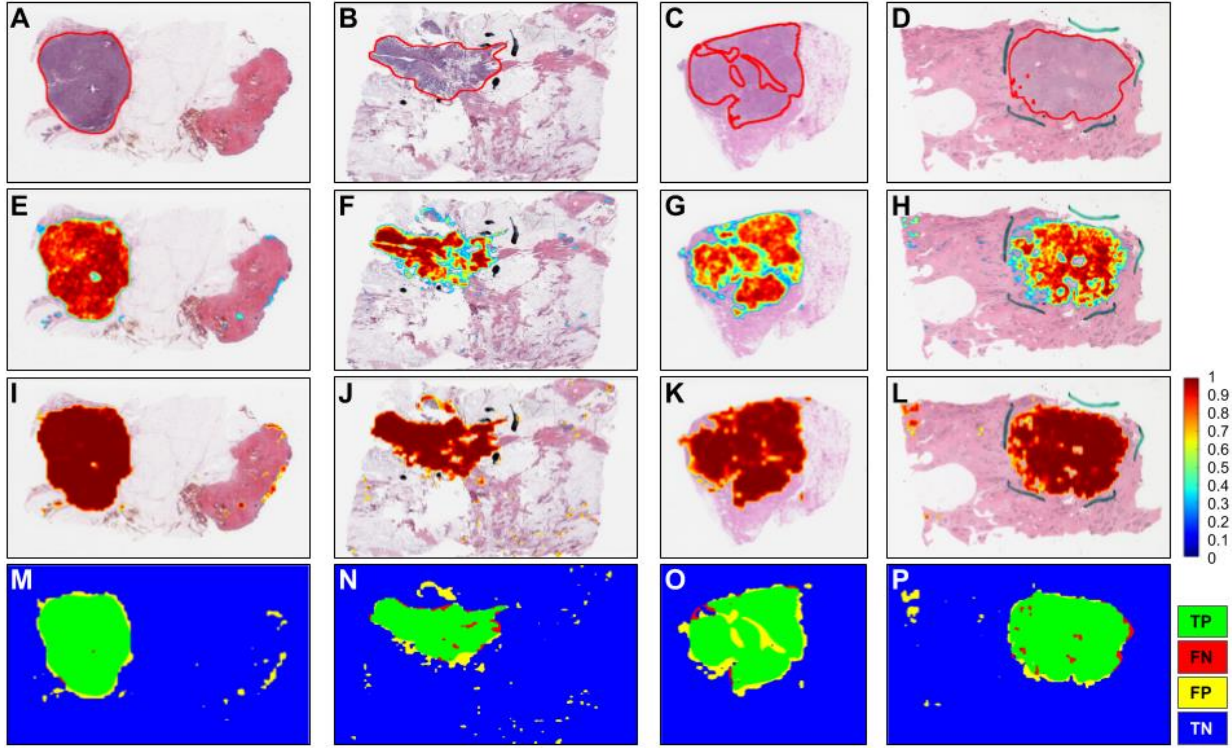

Fig. 3. Prediction map of representative slides from $T_{tcga}$. Figures A-D are WSIs with ground truth generated by our expert pathologists. Figures E-H show the corresponding prediction heatmap generated by our Cancer detection algorithm, C-Resnet34, prior to applying any aggregation methods. Figures I-L show the corresponding prediction map after applying Max aggregation function with window size of 4 then applying threshold of 0.6 to exclude prediction scores that are less than 0.6. Figures M-P show results of our algorithm in terms of TP (green), FN (red), FP (yellow), and TN (blue) regions.

| Method | F1-score | Accuracy | AUC |
|---|---|---|---|
| Saltz et al. [32] | 0.770 | 74.9% | 0.808 |
| L-VGG16 | 0.891 | 88.4% | 0.943 |
| L-Resnet34 | **0.893** | **89.0%** | **0.950** |
| L-Incepv4 | 0.879 | 87.5% | 0.938 |

TABLE 5
Performance comparison of the Lymphocytes detection task between Saltz et al. [32] and our models.

**Assessment of Inter-rater and Machine Versus Human Scoring of TIL Patches**

We performed a direct comparison of TIL predictions by the trained models with labeling of patches by experienced pathologists by scoring 8x8 "super-patches" for TIL content. Three pathologists assessed 500 super-patches as having low, medium, or high TIL content. Machine derived scores were assigned to a super-patch by counting TIL-positive patches in the super-patch; thus, the scores range from 0 to 64. To assess concordance between the human pathologists, we used the polychoric correlation coefficient, designed for comparing ordinal variables [52]. For comparison of continuous valued TIL counts estimated by the deep learning models versus those ordinal scores by the experienced pathologists (as having low, medium, or high TIL content), we used the polyserial coefficient. Table 6 shows the performance comparison between human raters with the models developed. We noticed a somewhat consistent improvement in the quality of concordance between human experts and machine-predictions, even perhaps slightly better than human-human concordance. Also, the deep learning models permit a lower variability relative to human raters, as evidenced by the width of the corresponding confidence intervals. Also the concordance between the summarized scores (using median) across pathologists vis-a-vis machine-derived predictions generally improves relative to concordance measures of individual experts against the machine predictions. As is seen in Figure 4, the median machine-derived score is quite distinct between the three ordinal bins.

| Rater | Human | VGG16 | Resnet34 | Incepv4 |
|---|---|---|---|---|
| A | R1:0.62[0.48,0.76] | 0.85 [0.81,0.88] | 0.82[0.78,0.86] | 0.85 [0.82,0.88] |
| R1 | R2:0.74[0.64,0.85] | 0.73[0.68,0.79] | 0.73[ 0.67,0.79] | 0.72 [0.66,0.78] |
| R2 | A:0.73 [0.62,0.84] | 0.73[0.68,0.79] | 0.74[0.69,0.80] | 0.76 [0.70,0.81] |
| Median | N/A | 0.77[0.70,0.83] | 0.74 [0.67,0.81] | 0.76[0.70,0.83] |

TABLE 6
Inter-rater concordance (between human raters: A, R1 and R2) and human vs. machine models. The point e of the correlation coefficients and confidence intervals are provided.

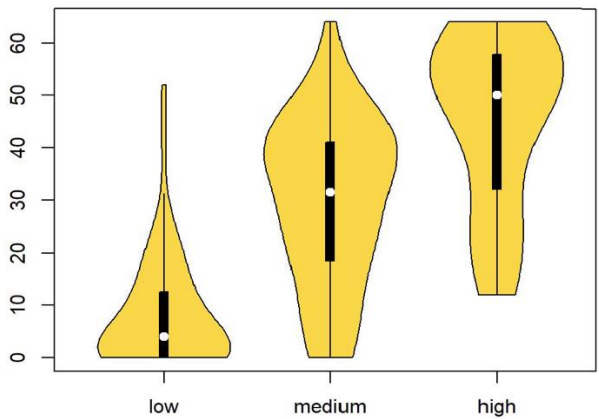

Fig. 4. Comparison of TIL scores of super-patches between pathologists and computational stain. x axis: median scores from three pathologists assessing 500 super-patches as having low, medium, or high lymphocyte infiltrate. y axis: scores from deep learning predictions on a scale from 0 to 64.

## DISCUSSION

Studies have shown tumor-infiltrating lymphocytes (TILs) can be used as a biomarker to predict clinical outcomes, including treatment response, in invasive breast cancer patients [9], [10], [11]. With the emergence of immunotherapy in breast cancer, the evaluation of the concentration of TILs as a readily available biomarker. As shown in Figure 3, the cancer detection algorithm shows that the cancer region occupies approximately 50-60% of the total tissue area in the WSI. The lymphocyte detection algorithm shows high probability areas with TILs. The tumor-TIL method provides insight about scattered TILs that occupy approximately 20-30% of the cancer region, consistent with a low TIL% categorization with additional spatial information that shows a sparse multi-focal distribution. Combined breast cancer tumor-TIL maps like the one shown in this example have been generated for 1,015 TCGA breast cancer WSIs and will be made publicly available in our custom web-based application.

The evaluation of TILs in invasive breast cancer is likely to expand due to the accumulating evidence showing how TILs can be used to predict treatment response in the settings of neoadjuvant and adjuvant chemotherapy. However, the routine evaluation of TILs has not achieved widespread adoption even though the established methodology by the International Immuno-Oncology Biomarker Working Group [18] is relatively straightforward, uncomplicated, and based on the examination of TILs on standard H&E-stained tissue sections. Figure 3 readily identifies TILs and a focal area with peritumoral TALs as a surrogate computational biomarker that is similar to how IHC is routinely utilized by pathologists to highlight cells and structures. However, IHC is not routinely performed to identify and classify subsets of TILs in breast cancer due to the time constraints of pathologists, desire to preserve diagnostic tissue, and additional costs, whereas this kind of insight can be made readily available in a low-cost and scalable manner to achieve the goals of the International Immuno-Oncology Biomarker Working Group. With emerging methods like our breast cancer tumor-TIL detection tool, pathologists will be able to add the evaluation of TILs to the standard IHC panel to determine ER, PR, HER2 expression status.

In previous work, several research groups carried out image analyses focused on detection of metastatic breast cancer [53], [54], [55] and mitosis [56], [57], [58] using highly curated but relatively small datasets from algorithm evaluation challenges [24], [25], [26], [27]. Cruz-Roa et al. 2017 and 2018 [28], [29] used deep learning approaches for detecting invasive breast cancer in WSIs. The deep learning models were trained using WSIs from the Hospital of the University of Pennsylvania (HUP) and from University Hospitals Case Medical Center/Case Western Reserve University (UHCMC/CWRU) and evaluated with 195 WSIs from TCGA. Kwok [30] and Dong et al. [31] proposed methods to classify breast cancer regions in WSIs using datasets provided by the ICIAR2018 Grand Challenge on Breast Cancer Histology Images [27]. The ICIAR2018 dataset contains 2 subsets of training data: Part A consists of 400 images of 2048x1536 pixels at 0.42 µm 0.42 µm resolution and Part B is made up 10 WSIs with manual annotations from pathologists. Kwok [30] implemented a 2-stage training approach where a basic CNN network is trained in the first stage to mine hard examples on data from part B. These examples were then used to train a deep learning model in the second stage. Dong et al. [31] employed deep reinforcement learning to decide whether regions of interest should be processed for segmentation at high or low image resolutions. Most recently, Amgad, M. et al., 2019 [59] proposed a fully convolutional framework for semantic segmentation of the first work using crowdsourcing in pathology task which involved a total of 25 participants at different expertise levels from medical students to expert pathologists to generate training data for a deep learning algorithm. The authors solely focused on segmenting triple-negative breast cancer (TNBC), an aggressive genomic subtype that comprises 15% of breast cancer cases, into five distinct classes: Tumor, Stroma, Inflammatory Infiltration, Necrosis and Other. Using a training dataset of 151 representative region of interests (ROI, mean ROI size of 1.18mm2) selected from 151 H&E TCGA WSIs with detailed curated annotations, a fully convolutional VGG16-FCN-8 network was able to achieve an AUC of 0.941 for Tumor region.

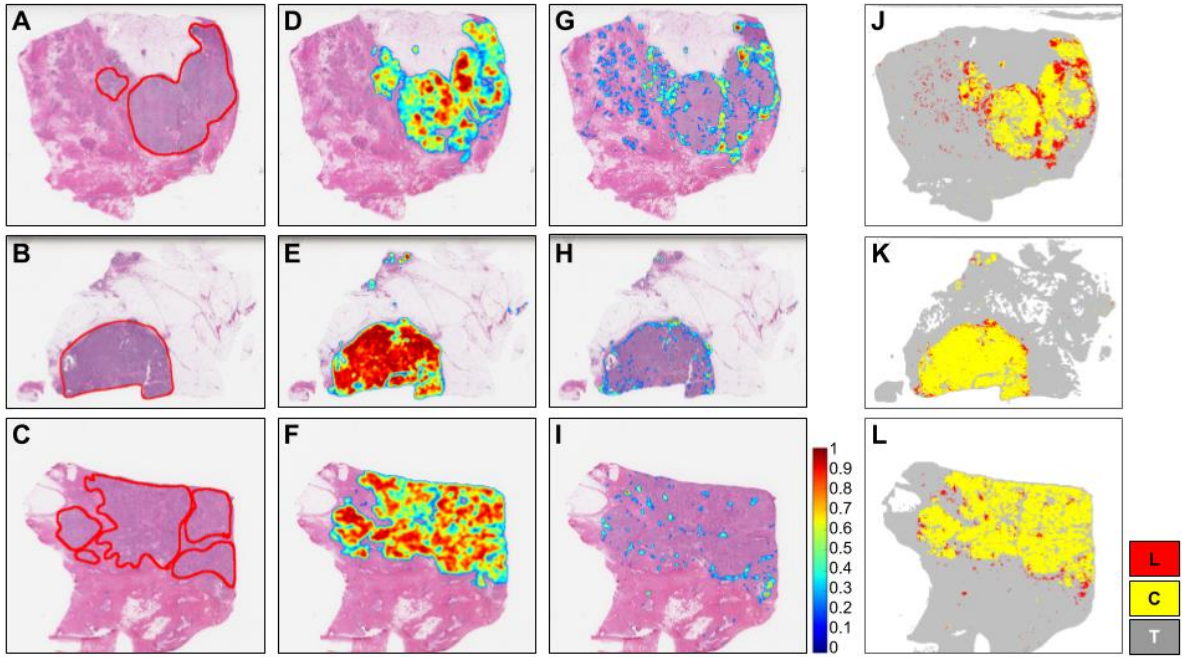

Fig. 5. Cancer and lymphocyte probability maps along with map of cancer and lymphocyte labels generated through analysis of representative slides from $T_{tcga}$ (A: TCGA-A2-A0CL-01Z-00-DX1, B: TCGA-A2-A04X-01Z-00-DX1, C: TCGA-A2-A0CW-01Z-00-DX1). Figures in a given row are results generated from the WSI depicted in the first column. Figures A-C depict WSIs with ground truth generated by our expert pathologist. Figures D-F depict the corresponding cancer probability maps generated by our cancer detection models, C-Resnet34. Figures G-I depict the corresponding lymphocyte probability maps generated by the Lymphocyte classification models, L-Resnet34. Figures J-L depict a combined heatmap of cancer and lymphocytes. Invasive breast cancer detection denoted in yellow with superimposed lymphocyte detection denoted in red. The legends of figures J-L are L, C, and T which refer to lymphocyte, cancer, and tissue region, respectively.

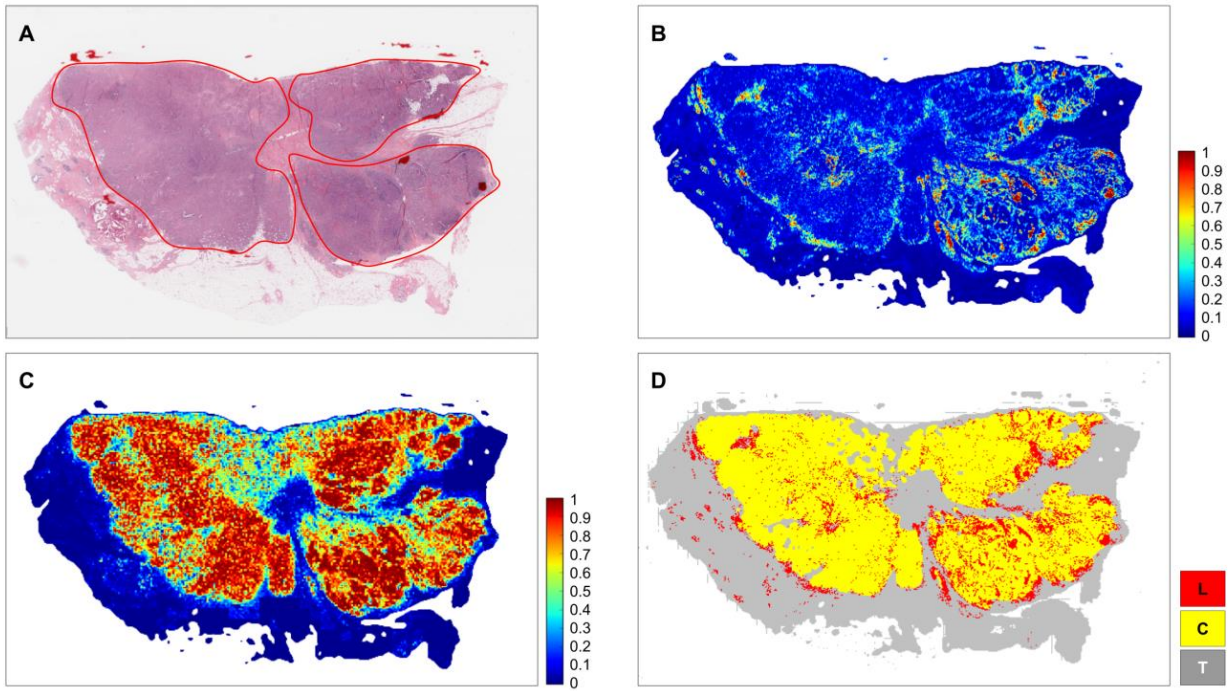

Fig. 6. Enlarged example of a cancer and lymphocyte probability map and cancer along with map of cancer and lymphocyte labels for TCGA WSI (case ID: TCGA-E9-A248-01Z-00-DX1) generated by our algorithms, C-Resnet34 and L-Resnet34. A: WSI of an invasive breast cancer H&E tissue section. The viable tumor region is annotated by a pathologist with a red line. B and C are lymphocyte probability map and cancer probability map predicted by our algorithm, respectively. The probabilities are in range from 0 to 1. D: Invasive breast cancer detection denoted in yellow with superimposed lymphocyte detection denoted in red. Grey areas outside of the yellow tumor region denote non-tumor connective and adipose tissues.

The current methods for assessing TILs in individual patients are still subjective, laborious, and may be difficult to quantify. More rigorous, objective, and efficient methods are needed. This is especially true for precision medicine applications since the tumor microenvironment in breast cancer is heterogeneous and composed of malignant cells, premalignant lesions, adjacent normal tissue, stroma, immune cell infiltrates, vessels, nerves, and fat. Therefore, to help further our understanding of breast cancer biology for research and clinical applications, we developed a tumor-TIL spatial mapping tool to automatically detect breast cancer in H&E stained WSIs to quantitatively estimate and characterize the relationship between tumor cells and TILs.

In the current state, the breast tumor-TIL maps can be used to identify spatial patterns of distributions of TILs within intra- and peritumoral regions of invasive cancer, as well as lymphocyte infiltrates in adjacent tissues beyond the borders of the tumor. This tool can also be adapted for practical uses that include improving the reproducibility and precision in reporting tumor size and features of the tumor boundary for radiologic-pathologic correlation. As a potential clinical application to quantify TILs and identify spatial patterns of distribution of TILs, this tool can help guide management and select treatment in conjunction with existing molecular subtyping platforms to predict survival and recurrence since TILs are being shown to be reliable prognostic and predictive biomarkers in invasive breast cancer. Another potential application of this tool is to screen candidates who may benefit from immunotherapy in primary, refractory, and recurrent disease since such treatments are not expected to be useful if a significant amount and distribution of TILs are not present.

Most existing software algorithms for TILs assessments are proprietary, expensive, and cannot be customized by the user. Therefore, we are making our invasive breast cancer TCGA tumor- TIL dataset publicly available with an interface to visually interact with the data. The interface permits quantification of TILs in tumor areas and the ability to rapidly spot check and evaluate true-positive and false-positive predictions by the deep learning models. The invasive breast cancer TCGA-TIL maps are displayed side-by-side with an interactive H&E slide viewer to permit a high level of exploration within the entire data set. We also intend to further combine this tumor-TIL method to characterize tumor immune heterogeneity and spatially characterize local patterns of the lymphocytic infiltrate in different parts of the tumor, e.g. center of the tumor, invasive margins, and metastases. The tumor- TIL heatmaps can also be combined with other types of digital pathology-based image analyses that extract object-level of information, such as size, shape, color, texture, etc. (collectively known as Pathomics), to generate an unprecedented quantitative examination of invasive breast cancer. Such analytic data can complement traditional histopathologic evaluation that can be correlated with clinical information, radiologic imaging, molecular studies, survival, and treatment response. We believe that the availability of Tumor-TIL maps along with software that allows interactive viewing of the computational analysis will improve reproducibility and precision in reporting tumor size, tumor boundary features, TILs assessment, and extraction of relevant nuclear and cellular features. These improvements will in turn enhance clinical and pathology decision support in guiding management, treatment selection, and predicting survival and recurrence, in conjunction with existing molecular subtyping platforms.

The need to quantify spatial inter-relationships between tumor regions and infiltrating lymphocytes is becoming increasingly important in invasive breast cancer. Tumor-TIL maps generated from H&E images can be employed to carry out a wide range of correlative studies in the context of clinical trials, epidemiological investigations, and surveillance studies. Our methods leverage open source convolutional neural networks; the programs we have developed are also being made public and freely available. In summary, our study has produced a reliable and robust methodology, datasets of TIL and cancer region predictions, and programs that can be employed to carry out tumor-TIL tissue image analyses of invasive breast cancers. In future studies, we will further refine our methodology and tools to differentiate between invasive and in situ premalignant lesions and explore methods that can facilitate faster predictions for practical real-time clinical applications.

**DATA AVAILABILITY**

The SEER images used in the training dataset were gathered in a work carried out with the SEER consortium and will be made publicly available in the future as part of a separate pilot project. The invasive breast cancer images are publicly available and provided by TCGA (http://cancergenome.nih.gov/ and the Genomic Data Commons (GDC) Data Portal in https://portal.gdc.cancer.gov/). The Cancer and TILs heatmaps for TCGA can be found at

https://stonybrookmedicine.box.com/v/tcga-brca-til-tumor-results

## ACKNOWLEDGEMENTS

This work was supported in part by 1U24CA180924-01A1, 3U24CA215109-02, and 1UG3CA225021-01 from the National Cancer Institute, R01LM011119-01 and R01LM009239 from the U.S. National Library of Medicine. This work used the Extreme Science and Engineering Discovery Environment (XSEDE), which is supported by National Science Foundation grant number ACI-1548562. Specifically, it used the Bridges system, which is supported by NSF award number ACI-1445606, at the Pittsburgh Supercomputing Center (PSC). NCI Surveillance Research Program overseeing the Virtual Tissue Repository (VTR) Pilot Program, from which participating SEER cancer registries (Greater California, Connecticut, Hawaii, Iowa, Kentucky, and Louisiana) supplied the whole slide images utilized in algorithm development and testing. The SEER VTR Pilot Program is supported by the Division of Cancer Control and Population Sciences at the National Cancer Institute of the National Institutes of Health.

## AUTHOR CONTRIBUTIONS

Conceptualization, J.S, T.K, R.G, H.L, A.L.V.D, D.S, D.F, T.Z, R.B; Methodology, J.S, T.K, H.L, A.R, L.H, S.A, D.S; Data Curation, R.G; Running Experiments, H.L, S.A; Writing – Original Draft, H.L, R.G, T.K, J.S.A, A.S; Writing – Review & Editing, H.L, R.G, T.K, J.S, A.R, A.L.V.D; Formal Analysis, J.S, R.G, A.R., A.L.V.D; Training Convolutional Neural Networks R.G, H.L; Supervision, J.S, T.K, D.S; Visualization, H.L, J.S.A, E.B; Software, T.K, E.B, J.S.A, A.S.